\documentclass[12pt]{iopart}  
\def\sq{\hbox {\rlap{$\sqcap$}$\sqcup$}}
\def\sq{\hbox {\rlap{$\sqcap$}$\sqcup$}}
\def\ried{$\widetilde{R}_{\ \ \mu\nu}^{ab}$}
\def\rie{$R_{\ \ \mu\nu}^{ab}$}
\def\ome{$\omega_{\ \ \mu}^{ab}$}
\def\omed{$\widetilde{\omega}_{\ \ \mu}^{ab}$}
\def\lsim{\raise0.3ex\hbox{$<$\kern-0.75em\raise-1.1ex\hbox{$\sim$}}}
\def\gsim{\raise0.3ex\hbox{$>$\kern-0.75em\raise-1.1ex\hbox{$\sim$}}}
\def\noi{\noindent}

\def\beq{\begin{equation}}   \def\eeq{\end{equation}}
\def\bea{\begin{eqnarray}}  \def\eea{\end{eqnarray}}
\def\nn{\nonumber}
\def\noi{\noindent}
\def\beeq{\begin{eqnarray}} \def\eeeq{\end{eqnarray}}
\newcommand\mysection{\setcounter{equation}{0}\section}
\renewcommand{\theequation}{\thesection.\arabic{equation}}
\newcounter{hran} \renewcommand{\thehran}{\thesection.\arabic{hran}}
\def\bmini{\setcounter{hran}{\value{equation}}
  \refstepcounter{hran}\setcounter{equation}{0}
  \renewcommand{\theequation}{\thehran\alph{equation}}\begin{eqnarray}}

\def\bminiG#1{\setcounter{hran}{\value{equation}}
\refstepcounter{hran}\setcounter{equation}{-1}
\renewcommand{\theequation}{\thehran\alph{equation}}
\refstepcounter{equation}\label{#1}\begin{eqnarray}}

\def\emini{\end{eqnarray}\relax\setcounter{equation}{\value{hran}}\renewcommand{\theequation}{\thesection.\arabic{equation}}}

\begin{document} 
\vbox to 1 truecm {}
\centerline{\Large\bf S-Dual Gravity in the Axial Gauge } 

\vskip 1 truecm

\centerline{\bf U. Ellwanger}
\centerline{Laboratoire de Physique Th\'eorique\footnote{Unit\'e Mixte
de Recherche (CNRS) UMR 8627}}  \centerline{Universit\'e de Paris XI,
B\^atiment 210, F-91405 ORSAY Cedex, France}
\vskip 2 truecm

\begin{abstract}
We investigate an action that includes simultaneously original and dual
gravitational fields (in the first order formalism), where the dual
fields are completely determined in terms of the original fields
through axial gauge conditions and partial (non-covariant) duality
constraints. We introduce two kinds of matter, one that couples to the
original metric, and dual matter that couples to the dual metric. The
linear response of both metrics to the corresponding stress energy
tensors coincides with Einstein's equations. In the presence of
nonvanishing standard and dual cosmological constants a stable solution
with a time independent dual scale factor exists that could possibly
solve the cosmological constant problem, provided our world is
identified with the dual sector of the model. \end{abstract}

\vskip 3 truecm

\noindent LPT Orsay 06-58\par
\noindent September 2006\par \vskip 3 truemm

\newpage
\pagestyle{plain}
\baselineskip 18pt

\mysection{Introduction}
\hspace*{\parindent}  The known concept of Hodge duality for form
fields (antisymmetric tensor fields), i.e. the permutation of free
equations of motion and Bianchi identities, can be generalized to
fields of mixed symmetry \cite{1r}-\cite{3r}  . Here again, linear
combinations of free equations of motion and Bianchi identities imply
Bianchi identities and free equations of motion for the dual fields or
their field strengths. In particular, this generalized Hodge duality
can be applied to linearized gravity in $d=4$ or higher dimensions
\cite{1r}-\cite{11r}.\par

One motivation for the introduction of a dual graviton are attempts to
realize hidden symmetries in $d=11$ supergravity/$M$-theory \cite{12r}.
However, only in $d=4$ the dual of a graviton is again a symmetric two
component tensor field \cite{1r}-\cite{9r}, that could be used to
construct a dual gravitational theory.\par

Any attempt in this direction has to face no-go theorems on local
Lorentz invariant dual models of gravity \cite{13r}-\cite{14r}. The way
out considered here is not to insist on a duality symmetry, and -- most
of all -- to break Lorentz symmetry twice: Once through a non-covariant
(axial) gauge fixing for the dual gauge fields, and through
non-covariant duality constraints on the field strengths.\par

Non-covariant gauges have the particular property that, in the context
of Yang-Mills theories, the gauge field can be reconstructed from the
field strength  \cite{15r,16r}. This feature has been used in
\cite{15r} in order to construct dual models of Yang-Mills theories
that are not, however, duality symmetric. \par

Here we study a similar program in the context of $4d$ gravity. In
order to formulate gravity as closely as possible to Yang-Mills
theories, we employ the first order Cartan formalism where vierbeins
and (spin) connections are treated as independent fields, and the
original local symmetries are diffeomorphisms and an ``internal''
$O(3,1)$ Lorentz symmetry (the gauge fields of the latter are the spin
connections).\par

Let us first outline the rough idea behind the present approach: the
starting point is a standard gravitational action involving vierbeins
$e^a_{\ \mu}$ and connections $\omega_{\ \ \mu}^{ab}$ (here and in the
following Latin indices indicate representations under the internal
Lorentz group, whereas Greek indices are space-time indices).\par

The Riemann-Cartan tensor \rie\ is the field strength of the
connections, and its dual \ried\  can be constructed by the contraction
of a pair of (antisymmetric) indices with two of the four indices of
the $4d$ completely antisymmetric epsilon tensor.\par

Once \ried\ is given through a duality relation in terms of \rie, one
can ask under which circumstances \ried\ can be written as a field
strength of a dual connection \omed. 

First, for each fixed choice of $[a\ b]$, \ried\ contains 6
antisymmetric combinations of indices $[\mu\ \nu]$. Deducting the gauge
degrees of freedom from \omed, we have however only three independent
connections \omed\ (for given $[a\ b]$) at our disposal. This leaves us
with two possibilities: (1) \ried\ has to satisfy certain constraints;
(2) we only use 3 out of the 6 antisymmetric combinations of indices
$[\mu\ \nu]$ in \ried\ in order to construct \omed.

The constraints (1) would be the (three independent) Bianchi
identities. In fact, in a weak field expansion the validity of the
Bianchi identities for \ried\ can be deduced from the ones for \rie\
under the condition that the Ricci tensor $R_{\mu \nu}$ vanishes 
\cite{1r}-\cite{9r}. Beyond a weak field expansion (where the
covariant derivatives acting on \ried\ would involve the not yet known
connection \omed, that differ in any case from \ome), or for
non-vanishing  $R_{\mu \nu}$, the Bianchi identities for \ried\ can no
longer be employed. We are left with the possibility (2) above: we
construct \omed\ from the 3 combinations of indices $[\mu\ \nu]$ that
are left over after a contraction of \ried\ with a fixed vector
$n^\nu$. Note that we still assume that all 6 combinations of indices
$[\mu\ \nu]$ of \ried\ are identified with the field strength of the
connection \omed; we do \underline{not} assume, however, that the 3
additional combinations of indices $[\mu\ \nu]$ of \ried\ are
determined through a duality relation in terms of \rie:

First, for a fixed vector $n^\mu$\footnote{In section 2
we actually employ a Lorentz vector $n^a$, but the essential idea is
more easily formulated with $n^{\mu}$.}, we define only 3
combinations of indices $[\mu\ \nu]$ of \ried\ in terms of \rie:
\beq
\label{1.1e}
n^{\mu} \widetilde{R}_{\ \ \mu\nu}^{ab} = {1 \over 2} \
\varepsilon^{ab}_{\ \ cd}\  n^{\mu} {R}_{\ \ \mu\nu}^{ab}\ .
\eeq

Next we introduce a dual connection $\widetilde{\omega}_{\ \
\mu}^{ab}$. On the dual connections we
impose the axial gauge condition 
 \beq
\label{1.2e}
\widetilde{\omega}_{\ \ \mu}^{ab} \ n^{\mu} = 0 \ .
\eeq

\noi Then we interpret \ried\ as a field strength for \omed, which
implies -- in the gauge (\ref{1.2e}) -- the relation
\beq
\label{1.3e}
n^{\mu} \partial_{\mu} \ \widetilde{\omega}_{\ \ \nu}^{ab} = n^{\mu}
\widetilde{R}_{\ \ \mu\nu}^{ab} \ .
\eeq

\noi Once $n^{\mu} \widetilde{R}_{\ \ \mu\nu}^{ab}$ is a known
expression in terms of \rie, the enforcement of both conditions
(\ref{1.2e}) and (\ref{1.3e}) determines the dual connections
completely (up to boundary conditions on a hypersurface perpendicular
to $n^{\mu}$) in terms of \rie.\par

A similar reasonning can be employed for the construction of a dual
vierbein $\widetilde{e}^a_{\ \mu}$, out of which a dual metric
$\widetilde{g}_{\mu\nu} = \widetilde{e}_{a\mu} \widetilde{e}_{\
\nu}^{a}$ can be constructed: For given \omed, the dual vierbein could
be constructed from the vanishing of the torsion tensor
\beq
\label{1.4e}
T_{\ \mu\nu}^a = \partial_{[\mu}\ \widetilde{e}_{\ \nu]}^a +
\widetilde{\omega}_{\ b[\mu}^a \ \widetilde{e}_{\ \nu]}^b \ .
\eeq
Again -- for each fixed value of $a$ -- $T_{\ \mu\nu}^a$ contains 6
antisymmetric combinations of indices $[\mu\ \nu]$, but (modulo gauge
degrees of freedom) we have only 3 degrees of freedom in 
$\widetilde{e}^a_{\ \mu}$ at our disposal. Again we can restrict
ourselves to the vanishing  of $T_{\ \mu\nu}^a n^\nu$, which allows to
construct the vierbeins $\widetilde{e}^a_{\ \mu}$ (in the axial gauge),
but the 3 extra components of $T_{\ \mu\nu}^a$ (for each $a$) will
vanish no longer in general. (One can prove that they would vanish in a
weak field expansion, but our aim is to go beyond it.)

Hence, we impose a gauge condition on $\widetilde{e}^a_{\ \mu}$
(i.e. a choice of the coordinate system) such that 
\beq
\label{1.5e}
\widetilde{e}^a_{\ \mu}\ n^{\mu} = \delta^a _{\ \mu}\ n^{\mu}
\eeq

\noi where $\delta^a _{\ \mu}$ is the Kronecker symbol. A coordinate
system where (\ref{1.5e}) holds can always be constructed locally.\par

Contracting (\ref{1.4e}) over $n^{\mu}$ and using both eqs.
(\ref{1.2e}) and  (\ref{1.5e}) one obtains
\beq
\label{1.6e}
n^{\mu}\partial_{\mu} \ \widetilde{e}_{\ \nu}^a =
\widetilde{\omega}_{\ b\nu}^a\ \delta_{\ \mu}^b n^{\mu}\ .
\eeq

\noi Similar to the case of the connection, eqs.  (\ref{1.5e}) and 
(\ref{1.6e}) determine the vierbeins $\widetilde{e}^a_{\ \mu}$
completely up to boundary conditions, now in terms of the connections
$\widetilde{\omega}^{ab}_{\ \ \mu}$ constructed before. Again, eqs.
(\ref{1.5e}) and  (\ref{1.6e})  do not imply the vanishing of all
components of $T_{\ \mu\nu}^a$ (again only half of them), i.e. the dual
theory is not entirely torsionless.\par

On the other hand we have now achieved our goal: the introduction and
definition of dual gravitational fields, including a dual metric
$\widetilde{g}_{\mu\nu}$, beyond a weak field expansion. Moreover, all
the previous conditions (\ref{1.1e}) -- (\ref{1.3e}), (\ref{1.5e}) and
(\ref{1.6e}) can be cast into the form of an action involving Lagrange
multipliers.\par

The possibility to construct an action for dual, and generally
interacting, fields is highly non-trivial: First, attempts to impose
(covariant) duality relations for all components of field strengths
using Lagrange multipliers leads generically to equations of motions
for these Lagrange multipliers that have non-trivial solutions
corresponding to new degrees of freedom. Attempts to eliminate all new
degrees of freedom through new multipliers result in the need for an
infinite tower of Lagrange multipliers \cite{17r}. The corresponding
problem can be cured quite easily here, with just one extra multiplier
required (see section 2). \par

Second, the complete permutation of non-linear equations of motion and
Bianchi identities (between original and dual fields) resulting from an
action is generally impossible, since an action allows for the addition
of sources for fields that modify the equations of motion, but leave
Bianchi identities unchanged. Here duality relations are satisfied by
only some of (half of) the field strengths -- that satisfy the second
Bianchi identities by construction -- and, furthermore, the first order
formalism allows for torsion that allows to violate the first Bianchi
(cyclic) identities of the Riemann tensor. \par

In fact, the possibility to construct an action involving
simultaneously the original and the dual gravitational fields as in the
following section 2 allows in particular to couple the dual metric, in
a standard fashion, to matter and to study its reactions to a stress
energy tensor. We will carry out this analysis in the weak field limit
in chapter 3, with the result that this reaction is the same as in
standard general relativity. This implies also that vacuum solutions of
standard general relativity remain valid also for the dual metric in
the weak field limit, but we have to expect modifications of the
solutions of the combined set of equations of motion (for both the
original and the dual fields) beyond lowest order. In particular any
attempt to integrate out completely the original gravitational fields
will result in non-local interactions for the dual fields, hence their
corresponding effective theory differs definitely from standard general
relativity.\par

The present approach is manifestly asymmetric between the original and
the dual gravitational fields; it seems generally impossible, however,
to implement a duality symmetry into interacting gravitational theories
\cite{13r,14r}.\par

In section 4 we study cosmological solutions of the equations of motion
derived in section 2. We consider simultaneous cosmological
constants in the standard and the dual sector and obtain a stable
solution with a static dual scale factor. This scenario could provide a
solution of the cosmological constant problem, provided our world is
identified with the dual sector of the model.\par

Finally, a summary and outlook is given in section 5.

\mysection{The action and its variation}
\hspace*{\parindent} 
The basic idea for the construction of an action
involving simultaneously original and dual gravitational fields has
been outlined in the introduction: Dual gravitational fields (the
connection and the vierbein) are introduced together with constraints
that determine them completely in terms of the original fields, up to
boundary conditions. This procedure guarantees the absence of new
degrees of freedom propagating in $4d$ space time, but the price to pay
is the explicit breaking of both internal (Lorentz) and space-time
(local coordinate reparametrizations) gauge symmetries. (The fixation
of space-time gauge symmetries could, in principle, be omitted, but
then the absence of new degrees of freedom is less obvious.)\par

In this section we present an action, and discuss the various steps of
symmetry breaking. In order to keep track of local coordinate
reparametrizations it turns out to be convenient to proceed slightly
differently than outlined in the introduction (although the final
result is the same): Instead of introducing a fixed constant vector
$n^{\mu}$, that would not be invariant under local coordinate
reparametrizations, we introduce a fixed constant Lorentz vector (but
space-time scalar) $n^a$ that is not invariant under local or global
Lorentz transformations. \par

We start with the standard Lagrangian for general relativity
in the first order Cartan formalism, where the independent fields are
the connection $\omega^{ab}_{\ \ \mu}$ and the vierbein $e_{\ \mu}^a$
(or its inverse $\overline{e}_a^{\ \mu}$). Lorentz indices $a$, $b$,
... are raised and lowered with the flat Lorentz metric
\beq
\label{2.1e}
\eta_{ab} = {\rm diag}\ (1, - 1, - 1, -1)\ .
\eeq

\noi The Riemann-Cartan curvature tensor is the field strength of the
connection:
\bea
\label{2.2e}
R_{\ \ \mu\nu}^{ab} &=& \partial_{\mu} \ \omega_{\ \ \nu}^{ab} -  
\partial_{\nu} \ \omega_{\ \ \mu}^{ab} + \omega_{\ c\mu}^{a}\ \omega_{\
\ \nu}^{cb} - \omega_{\ c\nu}^{a}\ \omega_{\ \ \mu}^{cb} \nn \\
&=& 2 \left (  \partial_{[\mu} \ \omega_{\ \ \nu ]}^{ab} + \omega_{\ c[
\mu}^{a}\ \omega_{\ \ \nu ]}^{cb}\right ) 
\eea

\noi The inverse vierbeins allow to construct a space-time scalar
version of the Riemann-Cartan curvature tensor, 
\beq
\label{2.3e}
R_{\ \ cd}^{ab} = \overline{e}_c^{\ \mu} \ \overline{e}_d^{\ \nu} \
R_{\ \ \mu\nu}^{ab}\ .
\eeq

\noi Then the standard Einstein Lagrangian is
\beq
\label{2.4e}
{\cal L}_E = {1 \over 2 \kappa} \det (e) \ R_{\ \ ab}^{ab}(e, \omega )\
.
\eeq

\noi (Our convention here and below is that all terms in the Lagrangian
transform as densities under local coordinate reparametrizations, such
that the final action is simply $\int d^4x {\cal L}$).\par

Next we introduce a new ``dual'' connection $\widetilde{\omega}_{\ \
\mu}^{ab}$, and a new ``dual'' vierbein $\widetilde{e}_{\ \mu}^a$ (or
its inverse $\widetilde{\overline{e}}_a^{\ \mu}$). At the first level
concerning $\widetilde{\omega}_{\ \ \mu}^{ab}$, we have to add three
Lagrange multiplier fields: \par

i) $K_{\ \ ab}^{(1)}$ will serve to impose the axial gauge condition on
$\widetilde{\omega}_{\ \ \mu}^{ab}$;\par

ii) $L_{\ \ ab}^{(1)\ c}$ will serve to impose that the dual of the
Riemann tensor (\ref{2.3e}), contracted with $n^d$ over its last index,
is equal to the field strength $\widetilde{R}^{ab}_{\ \ cd}$ of the
dual connection $\widetilde{\omega}_{\ \ \mu}^{ab}$ (defined in analogy
to (\ref{2.2e})), again contracted with $n^d$. \par

A few comments are in order here: In the case of standard duality
between (abelian) antisymmetric tensor fields, the dual field strength
is obtained by contracting an epsilon tensor (with space-time indices)
over the space-time indices of the original field strength, which would
correspond to the lower pair of indices of the Riemann tensors here.
Below we need, however, one of the lower indices -- contracted with
$n^d$ -- in order to employ a relation similar to eq. (\ref{1.3e}).
Therefore we employ the upper pair of indices in order to define a dual
Riemann tensor. In the absence of torsion, this would make no
difference, since then the Riemann tensor would be symmetric with
respect to an exchange of these two pairs of indices. Below we will
find, however, that torsion is generally non-vanishing here, i.e. not
all components of the connections $\omega^{ab}_{\ \ \mu}$ and
$\widetilde{\omega}_{\ \ \mu}^{ab}$ are related in the standard fashion
to the corresponding vierbeins. Hence the present definition of a dual
Riemann tensor in the sense of duality with respect to the internal
$O(3,1)$ Lorentz symmetry differs somewhat from standard
$S$-duality.\par

Finally we found it convenient to equate the Riemann tensors multiplied
with the determinants of the corresponding vierbeins, which simplifies
some of the equations of motion below and allows $L_{\ \ ab}^{(1)\ c}$
to transform as a space-time scalar.\par

iii) It turns out that the components of $L_{\ \ ab}^{(1)\ c}$ with the
index $c$ in the direction of $n^c$, i.e. $n_c L_{\ \ ab}^{(1)\ c}$,
are not constraint by the action. In order to eliminate these unwanted
degrees of freedom we employ another Lagrange multiplier $N^{(1)ab}$.
No further Lagrange multipliers are needed in order to cope with
unwanted degrees of freedom of $N^{(1)ab}$.\par

Hence at this first level, which treats the constraints on the dual
connection $\widetilde{\omega}_{\ \ \mu}^{ab}$, the following three
terms are added to the Lagrangian:
\bea
\label{2.5e}
&&{\cal L}^{(1)} = K_{\ \ ab}^{(1)}\ n^c \
\overline{\widetilde{e}}_c^{\ \mu}\ \widetilde{\omega}_{\ \
\mu}^{ab}\nn \\ 
&&+ L_{\ \ ab}^{(1)\ c}\ n^d \left \{ {1 \over 2} \det (e)
\varepsilon^{ab}_{\ \ ef} R_{\ \ cd}^{ef} (e, \omega ) - \det (
\widetilde{e}) \widetilde{R}_{\ \ cd}^{ab} (\widetilde{e}, 
\widetilde{\omega}) \right \} \nn \\
&&+ N^{(1)ab} \ n_c\ L_{\ \ ab}^{(1)\ \ c} \ .
\eea

\noi (Here $K^{(1)}$ and $N^{(1)}$ transform as densities like $\det
(\widetilde{e})$ under coordinate reparamet\-rizations). \par

For a given dual vierbein $\widetilde{e}_{\ \mu}^a$, all components of
$\widetilde{\omega}_{\ \ \mu}^{ab}$ are now fixed by the constraints
following from the variation of (\ref{2.5e}) with respect to $K_{\ \
ab}^{(1)}$ and $L_{\ \ ab}^{(1)\ c}$.\par

At the next level, we introduce constraints on the dual vierbein
$\widetilde{e}_a^{\ \mu}$ (or, for convenience, on its inverse
$\overline{\widetilde{e}}_a^{\ \mu}$). Again we will not follow exactly
the procedure outlined in the introduction (although the final result
will be the same), since we employ the constant Lorentz vector $n^a$
rather than a space-time vector $n^{\mu}$.\par

Otherwise, the r\^oles of the three Lagrange multipliers introduced at
this second level ressemble to the ones in (\ref{2.5e}):\par

i) $K^{(2)\ \nu}_{\ \ \mu}$ serves to impose an axial gauge condition
on $\widetilde{e}$;\par

ii) $L_{\ \ a}^{(2)\ b}$ serves to impose the vanishing of the torsion
tensor $T_{\ bc}^a$ contracted with $n^c$, with 
\beq
\label{2.6e}
T_{\ bc}^a = \overline{\widetilde{e}}_b^{\ \mu}\
\overline{\widetilde{e}}_c^{\ \nu}\ T_{\ \mu\nu}^a
\eeq

\noi and $T_{\ \mu\nu}^a$ as in (\ref{1.4e}). \par

iii) since the components $L_{\ \ a}^{(2)\ b} n_b$ are not determined
by varying the action, $N^{(2)a}$ serves to eliminate these degrees of
freedom.\par

Hence at the second level, the following three terms are added to the
Lagrangian:
\beq
\label{2.7e}
{\cal L}^{(2)} =K_{\mu}^{(2)\ \nu}\ n^b\ \partial_{\nu} \
\overline{\widetilde{e}}_b^{\ \mu} + L_{\ \ a}^{(2) \ b}\ T_{\
bc}^a n^c  + N^{(2)a}\ L_{\ \ a}^{(2) \ b} n_b \ . 
\eeq

\noi (Here $K^{(2)}$ and $L^{(2)}$ transform as densities under
coordinate reparametrizations).\par

Finally we can couple both kinds of gravitational fields to matter,
after defining the metrics
\bminiG{2.8e}
\label{2.8ae}
g_{\mu\nu} = e_{\mu a} \ e_{\nu}^{\ a}
\eeeq  
\beeq
 \label{2.8be}
\widetilde{g}_{\mu\nu} = \widetilde{e}_{\mu a} \ \widetilde{e}_{\nu}^{\
a} \ .
\emini 

For instance, one can introduce matter that couples to $g_{\mu\nu}$,
and another kind of matter that couples only to
$\widetilde{g}_{\mu\nu}$; hence two kinds of matter Lagrangians
\beq
\label{2.9e}
- {\cal L}_M (g) + \widetilde{\cal L}_M (\widetilde{g})\ .
\eeq

\noi (The relative minus sign in (\ref{2.9e}) serves just to reproduce
the standard signs in the corresponding Einstein equations, see section
3.)\par

Next we will derive the constraints imposed by the various Lagrange
multipliers.\par

Starting with ${\cal L}^{(1)}$ in (\ref{2.5e}), its variation with
respect to $K^{(1)}_{\ \ ab}$ imposes a gauge condition of the axial
type on $\widetilde{\omega}_{\ \ \mu}^{ab}$: 
\beq
\label{2.10e}
n^c \ \overline{\widetilde{e}}_c^{\ \mu} \ \widetilde{\omega}_{\ \
\mu}^{ab} \equiv n^{\mu} \ \widetilde{\omega}_{\ \ \mu}^{ab} = 0 \ .
\eeq

\noi Below we will find that for constant $n^a$, the vector 
\beq
\label{2.11e}
n^{\mu} = n^a\ \overline{\widetilde{e}}_a^{\ \mu}
\eeq

\noi is constant as well. \par

The variation of ${\cal L}^{(1)}$ with respect to $L^{(1)\ \ c}_{\ \
ab}$ includes, at first sight, a term proportional to $N^{(1)ab}$.
However, contracting the variation with $n^c$ and using the
antisymmetry of the curly bracket in (\ref{2.5e}) in $[cd]$, one
obtains 
\beq
\label{2.12e}
N^{(1)ab} = 0 \ .
\eeq

\noi Furthermore we can use that, with the definition analogously to
(\ref{2.2e}) and (\ref{2.3e}) of $\widetilde{R}_{\ \ cd}^{ab}$ and
(\ref{2.10e}),
\beq
\label{2.13e}
\widetilde{R}_{\ \ cd}^{ab}n^d = - \overline{\widetilde{e}}_c^{\
\mu}\ n^{\nu} \partial_{\nu} \ \widetilde{\omega}_{\ \ \mu}^{ab}
\eeq

\noi Hence one finally obtains the constraint from the variation w.r.t.
$L^{(1)\ \ c}_{\ \ ab}$
\beq
\label{2.14e}
\det (\widetilde{e})\ \overline{\widetilde{e}}_c^{\ \mu}\ n^{\nu} 
\partial_{\nu} \ \widetilde{\omega}_{\ \ \mu}^{ab} = - {1 \over 2} \det
(e) \varepsilon^{ab}_{\ \ ef}\ R^{ef}_{\ \ cd} (e, \omega )n^d \ .
\eeq

\noi The variation with respect to $N^{(1)ab}$ trivially implies
\beq
\label{2.15e}
L^{(1)\ \ c}_{\ \ ab}n_c = 0 \ .
\eeq

\noi Now we turn to ${\cal L}^{(2)}$ in (\ref{2.7e}), with the torsion
tensor $T^a_{\ bc} = T^a_{\ [bc]}$ defined in (\ref{2.6e}) and
(\ref{1.4e}). The variation with respect to $K^{(2)\ \nu}_{\ \ \mu}$
gives
\beq
\label{2.16e}
\partial_{\nu} \left ( n^b\ \overline{\widetilde{e}}_b^{\ \mu}\right )
\equiv \partial_{\nu}\ n^{\mu} = 0 \ .
\eeq

\noi Contracting the variation with respect to $L^{(2)\ b}_{\ \ a}$
with $n^b$ gives
\beq
\label{2.17e}
N^{(2) a} = 0 \ .
\eeq

\noi Using (\ref{2.17e}) and contracting the variation w.r.t. $L^{(2)\
b}_{\ \ a}$ with $\widetilde{e}^b_{\ \mu}$ gives 
\beq
\label{2.18e}
n^{\nu} \left ( \partial_{\mu}\ \widetilde{e}^a_{\ \nu} -
\partial_{\nu}\ \widetilde{e}^a_{\ \mu} +  \widetilde{\omega}_{\
b\mu}^{a}\ \widetilde{e}^b_{\ \nu} - \widetilde{\omega}_{\ b\nu}^{a}\
\widetilde{e}^b_{\ \mu}\right ) = 0 \ .
\eeq
\noi Using (\ref{2.16e}) and (\ref{2.11e}) one finds that the first
term in (\ref{2.18e}) vanishes, whereas (\ref{2.10e}) implies the
vanishing of the last term. Hence (\ref{2.18e}) collapses to 
\beq
\label{2.19e}
n^{\nu}  \partial_{\nu}\ \widetilde{e}^a_{\ \mu}  = n^b\ \widetilde{\omega}_{\ b\mu}^{a}\ .
\eeq

\noi Finally the variation with respect to $N^{(2) a}$ trivially
implies
\beq
\label{2.20e}
L^{(2)\ c}_{\ \ a} n_c = 0 \ .
\eeq

Let us study the consequences of the constraints (\ref{2.16e}) and
(\ref{2.19e}) on the dual metric $\widetilde{g}_{\mu \nu}$ in
(\ref{2.8be}). Contracting (\ref{2.19e}) with $n_a$ gives
\beq
\label{2.21e}
n^{\nu} \partial_{\nu}\left ( n_a \ \widetilde{e}^a_{\ \mu}\right ) =
0 \ .
\eeq

\noi From $n^{\mu}\widetilde{g}_{\mu \nu} = n_a  \widetilde{e}^a_{\
\mu}$ one finds that (\ref{2.21e}) is equivalent to 
\beq
\label{2.22e}
n^{\nu} \partial_{\nu}\left ( n^{\mu} \ \widetilde{g}_{\mu\nu}\right )
= 0 \ .
\eeq

\noi Axial gauges of the form (\ref{2.22e}) -- with constant $n^{\mu}$
as in (\ref{2.16e}) -- have been used in \cite{18r} in order to
construct graviton propagators and to analyze one loop diagrams in
quantized gravity. They can constrain the global properties of
space-time; the Schwarzschild solution in such a gauge -- with
$n^{\mu}$ time-like, $g_{00} = - 1$, $g_{0i} = 0$, i.e. in a comoving
frame (or Novikov coordinates) -- is known, however, and given e.g. in
\cite{19r}. In cosmology, on the other hand, this gauge is
standard.\par

Next we give the equations of motion that follow from the variations of
the complete action with respect to the vierbeins and the connections.
\par

First, from the variation with respect to $\overline{e}_a^{\ \mu}$,
contracted with $\overline{e}_g^{\ \mu}$ and after division by $\det
(e)$ one obtains
\bea
\label{2.23e}
&&{1 \over \kappa} \left [ - {1 \over 2}\ \delta_g^{\ a} \ R_{\ \
bc}^{bc} + R^{ab}_{\ \ gb}\right ] + \nonumber \\
&&L^{(1)\ \ [c}_{\ \ mn}\ n^{d]} \
\varepsilon^{mn}_{\ \ \ ef} \left [ - {1 \over 2}\ \delta_{\ g}^{a} \
R_{\ \ cd}^{ef} + \delta_{\ c}^{a} \ R^{ef}_{\ \ gd}\right ] - T_{M
g}^{\ \ \ a} = 0
\eea

\noi where
\beq
\label{2.24e}
T_{M g}^{\ \ \ a} = {\overline{e}_g^{\ \mu} \over \det ( e)} \ {\delta
\over \delta \overline{e}_a^{\ \mu}} \ {\cal L}_M\ .
\eeq

The variations with respect to $\omega^{ab}_{\ \ \mu}$ are best
expressed in terms of the two tensors
\beq
\label{2.25e}
E_{ab}^{\ \ \mu\nu} = E_{[ab]}^{\ \ \ [\mu \nu]} = \det (e) \
\overline{e}_a^{\ [\mu}\ \overline{e}_b^{\ \nu ]}
\eeq

\noi and
\beq
\label{2.26e}
S_{ab}^{\ \ \mu\nu} = S_{[ab]}^{\ \ \ [\mu \nu]} = \det (e) \  L^{(1)\
\ [c \ d]}_{\ \ ef\ \ n} \ \varepsilon_{\ \ ab}^{ef}\ 
\overline{e}_c^{\ \mu}\ \overline{e}_d^{\ \nu }\ .
\eeq

\noi Then one obtains
\beq
\label{2.27e}
{1 \over \kappa} \left [ \partial_{\nu} \ E_{ab}^{\ \ \mu\nu} - 2E_{
c[a}^{\ \ \ \mu\nu}\ \omega_{b]\  \ \nu}^{\ \ c}\right ] +
\partial_{\nu} \ S_{ab}^{\ \ \mu\nu} - 2S_{c[a}^{\ \ \ \mu\nu}\
\omega_{b]\  \ \nu}^{\ \ c} = 0\ .
\eeq

\noi For $S_{ab}^{\ \ \mu\nu} = 0$, eq. (\ref{2.27e}) would determine
$\omega^{ab}_{\ \ \mu}$ in terms of $e^a_{\ \mu}$, as usual, through
the vanishing of the covariant derivative of $e^a_{\ \mu}$, or the
vanishing of torsion. Generically, however, $S_{ab}^{\ \ \mu\nu}$ will
not vanish (since $L^{(1)\ \ c}_{\ \ e f}$ will not vanish), hence
torsion is generically non-vanishing for configurations $\omega^{ab}_{\
\ \mu}$ that solve eq. (\ref{2.27e}).\par

Concerning the variation with respect to $\overline{\widetilde{e}}_a^{\
\mu}$, it is convenient to consider the combination
\beq
\label{2.28e}
\overline{\widetilde{e}}_b^{\ \mu} \left ( \delta^a_{\ d} - {n^a n_d
\over n^2}\right ) {\delta \over \delta \overline{\widetilde{e}}_a^{\
\mu}}
\eeq

\noi that is independent of the Lagrange multiplier field $K^{(2)\
\nu}_{\ \ \mu}$ in (\ref{2.7e}). (The remaining components just
determine $K^{(2)\ \nu}_{\ \ \mu}$, that appears nowhere else.) Then
one finds
\bea
\label{2.29e}
&&\left ( \delta_a^{\ b} - {n_an^b \over n^2} \right ) \Big \{ 2 L^{(1)\
\ c}_{\ \ ef} \delta_{[b}^{\ \ d} \ \widetilde{R}^{ef}_{\ \ c]g}n^g
\nonumber \\
&&
- {1 \over 2} \ \widetilde{e}^d_{\ \mu} \ n^g \
\overline{\widetilde{e}}_g^{\ \nu} \ \partial_{\nu} \left ( L^{(2)\
c}_{\ \ b}\ \overline{\widetilde{e}}_c^{\ \mu}\right ) +
\widetilde{T}_{M b}^{\ \ \ d} \Big \} = 0
\eea

\noi with
\beq
\label{2.30e}
 \widetilde{T}_{M b}^{\ \ \ d} = {\overline{\widetilde{e}}_b^{\ \mu}
\over \det (\widetilde{e})}\ {\delta \over \delta
\overline{\widetilde{e}}_d^{\ \mu}}\ \widetilde{\cal L}_M \ .
\eeq

Finally the variation with respect to $\widetilde{\omega}^{ab}_{\ \
\mu}$ gives, again after an elimination of $K^{(1)}_{\ \ ab}$ through a
contraction with $n^d \ \overline{\widetilde{e}}_d^{\ \mu}$,
\beq
\label{2.31e}
n^d\ \overline{\widetilde{e}}_d^{\ \nu}\ \partial_{\nu} \left ( \det
(\widetilde{e} )\ L_{\ \ ab}^{(1) \ c}\  \overline{\widetilde{e}}_c^{\
\mu} \right )  - {1 \over 2} L^{(2)\ \ c}_{\ \ [ a}\ {n}_{b]}
\ \overline{\widetilde{e}}_c^{\ \mu} = 0 \ .
\eeq 

\noi Some consequences of these equations of motions will be studied in
the next sections. \par

Before concluding this section we have to make some comments on the
Lorentz symmetry breaking induced by the constant Lorentz vector $n^a$.
Apart from the terms that impose axial gauge conditions (the first
terms in (\ref{2.5e}) and (\ref{2.7e})), $n^a$ appears in various other
terms in (\ref{2.5e}) and (\ref{2.7e}). In the next section we study
the equations of motion in a weak field expansion (around a Minkowski
vacuum, and considering all Lagrange multipliers as weak fields). At
this level we find no explicit Lorentz symmetry breaking in the
equations of motion.\par

Gravitational self-interactions, that start to play a role in higher
order in a weak field expansion, will generally not respect Lorentz
invariance, however. In the worst case this could induce violations of
unitarity and/or strong gravitational self-couplings at inacceptable
length scales. \par

It may then be advisable to promote $n^a$ to a field $n^a(x)$, and to
replace explicit Lorentz symmetry breaking by spontaneous Lorentz
symmetry breaking via a potential $\lambda (n^a n_a \pm 1)^2$ in the
Lagrangian for $n^a (x)$. Apart from a massive radial mode, this
scenario will add three Nambu-Goldstone (NG) modes to the model.\par

The possible fate of such NG modes has recently been reviewed in
\cite{20r}, and depends heavily on the Lagrangian (e.g. additional
$R^2$ terms, that generate a propagating spin connection): the NG
modes could remain massless, but could also become massive degrees of
freedom. Some phenomenological consequences of such NG modes have been
studied in \cite{21r}, but even the quadratic part of an effective
Lagrangian would have been to be constructed here. 


Also, unless the Lorentz gauge symmetry is restored this way, the axial
gauge condition (2.5) imposed on $\widetilde{\omega}^{ab}_{\ \ \mu}$
are not just gauge conditions that can, in principle, be chosen at
will (in contrast to the gauge conditions (2.7) on
$\widetilde{e}^a_\mu$): observables will depend on the present choice
of the gauge condition imposed on $\widetilde{\omega}^{ab}_{\ \ \mu}$
-- however, only to higher order in a weak field expansion, see the
next chapter.

\mysection{The linearized gravitational field equations}
\hspace*{\parindent}  The aim of this section is to study the
linearized response of the gravitational fields to matter sources,
notably to a dual stress energy tensor $\widetilde{T}_{M a}^{\ \ \ b}
(\widetilde{g})$. \par

The point is that, as long as $T_{M a}^{\ \ \ b}(g)$ does not depend on
$\widetilde{g}$ and vice-versa (what we will assume in the following),
standard matter described by $T_{M a}^{\ \ \ b}(g)$ ``sees'' a
space-time geometry described by $g_{\mu\nu}$, whereas dual matter
described by $\widetilde{T}_{M a}^{\ \ \ b} (\widetilde{g})$ ``sees'' a
space-time geometry described by $\widetilde{g}_{\mu\nu}$ that will
generally be quite different! An interesting question is then whether
$\widetilde{g}_{\mu\nu}$ reacts to $\widetilde{T}_{M a}^{\ \ \ b} $ in
a way that resembles the standard Einstein equations; if this is
the case, our world could possibly correspond to the dual matter of the
model.\par

An answer to this question is not quite trivial even to lowest order,
where (counting Lagrange multipliers as weak fields) terms $\sim L
\cdot R$ in eqs. (\ref{2.23e}) and (\ref{2.29e}) can be neglected:
whereas the effect of $T_{M a}^{\ \ \ b}$ on $g_{\mu\nu}$ is obvious
from (\ref{2.23e}) (and agrees with Einstein's gravity), the effect of
$\widetilde{T}_{M a}^{\ \ \ b} $  is\par

i) a non-vanishing value of $L^{(2)\ c}_{\ \ b}$ from (\ref{2.29e}),
which induces\par

ii) a non-vanishing value of $L^{(1)\ \ c}_{\ \ ab}$ from
(\ref{2.31e}), which induces\par

iii) non-vanishing torsion for $\omega^{ab}_{\ \ \mu}$ from
(\ref{2.27e}), hence non-vanishing components of $R^{ab}_{\ \ \mu\nu}$,
which induce\par

iv) a non-vanishing $\widetilde{\omega}^{ab}_{\ \ \mu}$ from
(\ref{2.14e}), that finally generates\par

v) a non-vanishing $\widetilde{e}^a_{\ \mu}$ from (\ref{2.19e}).\par

Subsequently we will carry out these steps explicitly. First it is
convenient, however, to adopt a convention concerning the direction of
$n^a$, that we assume to be time-like: 
\beq
\label{3.1e}
n^a = (1, 0, 0, 0)
\eeq

\noi and the first component will be denoted by 0, i.e. $n^0 = 1$. All
space and Lorentz coordinates perpendicular to $n^{\mu} \sim n ^a$ (to
lowest order) will be denoted by latin letters $i, j, k, \dots$ from
the middle of the alphabet.\par

Next we have to comment the fact that eq. (\ref{2.29e}) contains a
projector such that the components $\widetilde{T}_{M 0}^{\ \ \ d}$ do
not appear. Such a situation is actually familiar from Yang-Mills
theories in axial gauges: After imposing the axial gauge condition, the
corresponding components of the currents (here: the stress energy
tensor) decouple from the gauge field. They contribute nevertheless to
the dynamics of the theory through the equations associated with
(covariant) current conservation, that can be derived via Noether's
theorem. Likewise, we have to use the covariant conservation of
$\widetilde{T}_{M a}^{\ \ \ b}$ (and its symmetry) here, which gives to
lowest order 
\beq
\label{3.2e}
\partial_0 \ \widetilde{T}_{M a}^{\ \ \ 0} + \partial_i \
\widetilde{T}_{M a}^{\ \ \ i} = 0 
\eeq

\noi for all $a$. Once $\widetilde{T}$ is assumed to be symmetric, eq.
(\ref{3.2e}) determines all components of $\widetilde{T}_{M a}^{\ \ \
b}$ in terms of $\widetilde{T}_{M i}^{\ \ \ j}$, hence the absence of
$\widetilde{T}_{M 0}^{\ \ \ d}$ in (\ref{2.29e}) constitutes no longer
a paradox.\par

Now we turn to eq. (\ref{2.29e}) which reads to lowest order (where
$\widetilde{e}^d_{\ \mu} \sim \delta^d_{\ \mu}$,
$\overline{\widetilde{e}}_g^{\ \nu} \sim \delta_g^{\ \nu}$, and only $b
= j$ contributes)
\beq
\label{3.3e}
\partial_0 \ L^{(2)\ i}_{\ \ j} = 2  \widetilde{T}_{M j}^{\ \ \ i} \ . 
\eeq

 The index combination $(ab) = (j0)$ of eq.  (\ref{2.31e}) gives, with 
(\ref{3.3e})  for $L^{(2)}$,
\beq
\label{3.4e}
\partial_0 \ \partial_0\ L^{(1)\ \ i}_{\ \ j0} = {1 \over 2}\  
\widetilde{T}_{M j}^{\ \ \ i} \ , 
\eeq

\noi whereas the combination $(ab ) = (jk)$ implies the vanishing of
$L^{(1)\ \ i}_{\ \ jk}$ up to terms linear in $t$. Eq.  (\ref{3.4e})
implies a non-vanishing value for $S_{ab}^{\ \ \mu\nu}$ in 
(\ref{2.26e}), which gives
\beq
\label{3.5e}
S_{ij}^{\ \ k0} = \varepsilon_{\ \ ij}^{l0}\ L^{(1)\ \ k}_{\ \ l0} =
\varepsilon_{\ \ ij}^{l0}\ {1 \over 2 \partial_0 \partial_0} \
\widetilde{T}_{M l}^{\ \ \ k}
 \eeq

\noi where we allowed ourselves to represent integrals with respect to
$x^0 = t$ by an inverse derivative $1/\partial_0$. Next we have to
determine $\omega^{ab}_{\ \ \mu}$ from (\ref{2.27e}), and it is
convenient to decompose $\omega_{ab\mu}$ into 
\beq
\label{3.6e}
\omega_{ab\mu} = \Omega_{ab\mu}(e) + \omega_{ab\mu}^T
\eeq
\noi where
\beq
\label{3.7e}
\Omega_{ab\mu} (e) = {1 \over 2} \left ( \partial_b \left ( e_{a\mu} +
e_{\mu a}\right ) - \partial_a \left ( e_{b \mu} + e_{\mu b}\right ) +
\partial_{\mu} \left ( e_{ba} - e_{ab}\right ) \right ) 
\eeq

\noi and $\omega_{ab\mu}^T$ represents torsion. If one would replace
$\omega$ by $\Omega$ in (\ref{2.27e}), its first line would vanish
identically. Hence (\ref{2.27e}) determines $\omega^T$ in terms of
$S_{ab}^{\ \ \mu\nu}$, and $\Omega$ (or $e$) has to be determined
elsewhere (by eq. (\ref{2.23e})). Exploiting the various index
combinations of eq. (\ref{2.27e}) one obtains after some calculation
\bea
\label{3.8e}
&&\omega^T_{\ ijk} = - {\kappa \over 2 \partial_0} \left (
\varepsilon^{n0}_{\ \ ik} \ \widetilde{T}_{M jn} +
\varepsilon^{n0}_{\ \ ij} \ \widetilde{T}_{M kn} +
\widetilde{\varepsilon}^{n0}_{\ \ kj} \ \widetilde{T}_{M in}  \right
)\ ,\nn \\
&&\omega^T_{\ i0j} = - {\kappa \over 2 \partial_0}\ \varepsilon^{n0}_{\
\ ij} \ \widetilde{T}_{M 0n} \ , \nn \\ 
&&\omega^T_{\ ij0} = - {\kappa \over 2 \partial_0}\ \varepsilon^{n0}_{\
\ ij} \ \widetilde{T}_{M 0n} \ .
\eea

Next we wish to construct $\widetilde{R}_{\ \ cd} ^{ab}
(\widetilde{\omega})$ from $R^{ab}_{\ \ cd}(\omega )$ via (\ref{2.14e})
or, better, directly from the constraint from (\ref{2.5e}):
\beq
\label{3.9e}
\widetilde{R}_{\ \ cd} ^{ab} = {1 \over 2} \ \varepsilon^{ab}_{\ \ ef}\
R^{ef}_{\ \ cd} \ .
\eeq

At first sight we have a problem, since $R^{ef}_{\ \ cd}$ depends on
$\omega = \Omega + \omega^T$, and $\Omega$ is not yet known. However,
it turns out that the linearized dual Ricci tensor does not depend on
$\Omega$, since  $\varepsilon^{ab}_{\ \ ef} R^{ef}_{\ \ ad}(\Omega ) =
0$ identically. This allows  us to construct 
\beq
\label{3.10e}
\widetilde{R}_{\  c} ^{a} = \widetilde{R}_{\ \ cb} ^{ab} = {1\over 2}
\varepsilon^{ab}_{\ \ ef} R^{ef}_{\ \ cb}(\omega^T )
\eeq

\noi with (\ref{3.8e}) for $\omega^T$, and using (\ref{3.2e}) in order
to rewrite spacial derivatives. The result is 
\beq
\label{3.11e}
\widetilde{R}_{\  b} ^{a} = \kappa \left ( \widetilde{T}_{M \ \ b}
^{\ \ a} - {1 \over 2}\ \delta^a_{\ b} \ \widetilde{T}_{M \ \ c} ^{\ \ c
}  \right ) 
\eeq

\noi which seems to coincide with the standard Einstein equations.\par

However, $\widetilde{R}_{\ \ \mu\nu} ^{ab}$ is the Riemann-Cartan
tensor defined in terms of $\widetilde{\omega}_{\  \ \mu} ^{ab}$, and
coincides with the Riemann tensor $\widetilde{R}_{\ \ \mu\nu}
^{ab}(\widetilde{e})$ only if the dual connection
$\widetilde{\omega}_{\ \ \mu} ^{ab}$ is torsionless, i.e. if 
\beq
\label{3.12e}
\widetilde{\omega}_{\  \ \mu} ^{ab} = \widetilde{\Omega}_{\  \ \mu}
^{ab}(\widetilde{e})
\eeq

\noi with $\widetilde{\Omega}(\widetilde{e})$ as in (\ref{3.7e}). In
order to study this question, we have to construct all components of
$\widetilde{\omega}_{\  \ \mu} ^{ab}$ from eq. (\ref{2.14e}), and
subsequently $\widetilde{e}^a_{\ \mu}$ from (\ref{2.19e}). The result
is that (\ref{3.12e}) holds indeed, provided \par

i) the original Ricci tensor $R^a_{\ b} (\omega = \Omega + \omega^T)$
satisfies
\bea
\label{3.13e}
&&R^0_{\ i} = 0 \ , \nn \\
&&\partial_0\ R^{i}_{\ j} = 0\ ;
\eea

ii) the off-diagonal components $\widetilde{e}^0_{\ i}$ of the dual
vierbein, that are required to be $t$-independent from eq.
(\ref{2.21e}), satisfy
\beq
\label{3.14e}
\partial_{[ i}\ \widetilde{e}^0_{\ j]} = 0 \ .
\eeq

\noi (Actually, we could have imposed $n_a \widetilde{e}^a_{\ \mu} =
n_a \delta^a_{\ \mu}$, i.e. $\widetilde{e}^0_{\ j} = 0$, from the
beginning).\par

Equations (\ref{3.13e}) follow from the not yet considered eq.
(\ref{2.23e}) if $T_{M g}^{\ \ \ a} = 0$, but they also allow for a
cosmological constant $T_{M g}^{\ \ \ a} = \delta_g^{\ a}\Lambda$ in
the ``standard'' matter Lagrangian.\par

Note finally that for a vanishing dual stress energy tensor, the
equations (\ref{3.8e}) above imply vanishing torsion for the standard
spin connection, and the linearized eqs. (\ref{2.23e}) can be
interpreted as standard torsionless Einstein equations. \par

To summarize, we have learned about two important features of the
present model in this chapter: \par

i) the possibility to reproduce the linearized Einstein equations
including matter for the dual gravitational fields, in spite of the
Lorentz-non-covariant action, \par

ii) the crucial role played by torsion within the present first order
formalism: ``standard'' torsion is induced by ``dual'' matter (and
vice versa); this allows to generalize the known correspondence
between vacuum equations of motion and Bianchi identities
\cite{1r}-\cite{7r} to equations of motion with sources, a possibility
already advocated (in a cosmological context) in \cite{22r}.

\mysection{Cosmological solutions}
\hspace*{\parindent} In this section we return to the full nonlinear
equations of motion of section 2 and study cosmological solutions. The
aim is to check under which circumstances the standard
Freedman-Robertson-Walker (FRW) equations are reproduced, but also to
see whether the model could provide a hint for a solution of the
cosmological constant problem (CCP). A solution to the CCP would
correspond to a (stable) solution of the equations of motion where
$\Lambda$ and/or $\widetilde{\Lambda}$ are non-vanishing, but the
standard and/or dual metric remains nearly time-independent (does not
explode exponentially with cosmic time). Indeed we will find such
solutions for the dual metric, with non-vanishing  $\Lambda$ and
$\widetilde{\Lambda}$, below. \par

First we have to make a general ansatz for all fields and Lagrange
multipliers of the model, that is consistent with a homogeneous and
isotropic universe (i.e. depend on $x^0 = t$ only). For the Lorentz
vector $n^a$ we will make the same choice as in eq. (\ref{3.1e}). Note
that we do not have enough gauge symmetries in order to gauge the
component $e^0_{\ 0}$ of the standard vierbein to 1, whereas the
component $\widetilde{e}^0_{\ 0}$ of the dual vierbein is constraint to
be constant (that can be chosen as 1) by eq. (\ref{2.21e}). Then the
most general ansatz is as follows (where $a$, $b$, $r$, $s$,
$\widetilde{a}$, $\widetilde{r}$, $\widetilde{s}$, $\ell^{(i)}$ are
functions of $t$):
 \bea
\label{4.1e}
&&e^a_{\ \mu} : = {\rm diag} (b, a, a, a)\nn \\
&&\omega^{ab}_{\ \ \mu} : \omega^{0i}_{\ \ j} = r \delta^i_{\ j},\
\omega^{ij}_{\ \ k} = s \varepsilon^{ij}_{\ \ k}\nn \\
&&\widetilde{e}^a_{\ \mu} : = {\rm diag} (1, \widetilde{a},
\widetilde{a}, \widetilde{a})\nn \\
&&\widetilde{\omega}^{ab}_{\ \ \mu} : \widetilde{\omega}^{0i}_{\ \ j} =
\widetilde{r} \delta^i_{\ j},\ \widetilde{\omega}^{ij}_{\ \ k}
= \widetilde{s} \varepsilon^{ij}_{\ \ k} \nn \\
&&L^{(1)\ \ c}_{\ \ ab} : L^{(1)\ \ j}_{\ \ i0} = \ell^{(1)}
\delta_i^{\ j},\ L^{(1)\ \ k}_{\ \ ij} = \overline{\ell}^{(1)}
\varepsilon_{ij}^{\ \ k} \nn \\
&&L^{(2)\ b}_{\ \ a} : L^{(2) \ j}_{\ \ i} = \ell^{(2)} \delta_i^{\ j}
\eea

\noi (concerning the Lagrange multipliers, we have taken care of the
constraints (\ref{2.15e}) and  (\ref{2.20e})).\par

Plugging these ans\"atze into the equations (and constraints) of
section 2, we obtain from eq.  (\ref{2.19e}) (where dots denote time
derivatives):
\beq
\label{4.2e}
\dot{\widetilde{a}} = - \widetilde{r}\ , 
\eeq

\noi from (\ref{2.14e}) with $(abc) = (i0j)$ and $(ijk)$: 
\bminiG{4.3e}
\label{4.3ae}
\widetilde{a}^2  \  \dot{\widetilde{r}} = - a^2 \dot{s}\ , 
\eeeq  
\beeq
\label{4.3be}
\widetilde{a}^2  \  \dot{\widetilde{s}} = - a^2 \dot{r}\ , 
\emini

\noi from (\ref{2.31e}) with $(ab) = (i0)$ and $(ij)$:
\bminiG{4.4e}
\label{4.4ae}
\ell^{(2)} = 4\widetilde{a}\ \partial_0 \left ( \ell^{(1)} \
\widetilde{a}^2 \right ) \ , 
\eeeq  
\beeq
\label{4.4be}
2\widetilde{a}\ \partial_0 \left ( \overline{\ell}^{(1)} \
\widetilde{a}^2 \right ) = 0 \ .
\emini

\noi These equations are used to simplify some of the equations below,
notably to eliminate $\ell^{(2)}$. \par

For the stress energy tensor $T_{M}$ we
will make, to start with, the general ans\"atze
\beq
\label{4.5e}
T_{M 0}^{\ \ \ 0} ,\ T_{M i}^{\ \ \ j} =
\delta_i^{\ j} \ T_{M , S} \ , 
\eeq
\noi and for $\widetilde{T}_{M}$ 
\beq
\widetilde{T}_{M i}^{\ \ \ j}  =
\delta_i^{\ j} \ \widetilde{T}_{M , S} \ .
\label{4.6e}
\eeq

Note that $\widetilde{T}_{M 0}^{\ \ \ 0}$ does not contribute to eq.
(\ref{2.29e}). As stated before, this does not constitute a paradox,
since $\widetilde{T}_{M 0}^{\ \ \ 0}$ has to be determined by the
conservation law that assumes, in the present context, the form 
\beq
\label{4.7e}
\dot{\widetilde{T}}_{M 0 0} + 3{\dot{\widetilde{a}} \over
\widetilde{a}} \left (  \widetilde{T}_{M 00} - \widetilde{T}_{M,
S}\right ) = 0 \ .
\eeq

\noi Next eq. (\ref{2.23e}) gives, for $(ag) = (00)$ and $(ij)$, 
\bminiG{4.8e}
\label{4.8ae}
3\left ( r^2 - s^2\right ) = \kappa \ a^2 \ T_{M 00}\ , 
\eeeq  
\beeq
\label{4.8be}
-2 \dot{r} + {b \over a} \left ( r^2 - s^2\right ) + 4 \kappa
\left ( \ell^{(1)}\dot{s} - \overline{\ell}^{(1)} \dot{r} \right ) =
\kappa \ ab\  T_{M, S}\ .
\emini

\noi Eq. (\ref{2.27e}) gives, for $(ab\mu ) = (0ij)$ and $(ijk)$ (the
other index combinations just give $0 = 0$):
\bminiG{4.9e}
\label{4.9ae}
r + {\dot{a} \over b} = - {\kappa \over ab} \ \partial_0 \left (
\overline{\ell}^{(1)} a^2 \right ) \  ,
\eeeq  
\beeq
\label{4.9be}
s = {\kappa \over ab} \   \partial_0 \left ( \ell^{(1)}  a^2 \right  )
\  .
\emini

\noi Finally eq. (\ref{2.29e}) gives
\beq
\label{4.10e}
4\widetilde{a} \left ( \ell^{(1)}\dot{\widetilde{r}} -
\overline{\ell}^{(1)} \dot{\widetilde{s}}\right ) - 2 \partial_0
\partial_0  \left ( \ell^{(1)}\widetilde{a}^2 \right ) = -
\widetilde{a}^2 \widetilde{T}_{M , S}\ .
\eeq

No further equations can be derived, and we are left with indeed 10
equations (\ref{4.2e})-(\ref{4.4e}), (\ref{4.8e})-(\ref{4.10e}) for 10
functions in the ansatz (\ref{4.1e}).\par

First it can be checked that, in the absence of dual matter
($\widetilde{T}_{M , S} = 0$ in eq. (\ref{4.10e})), we can put $s=
\ell^{(1)} = \overline{\ell}^{(1)}  = 0$ (and $\widetilde{a} =$
const.), and equations (\ref{4.8e}) collapse to (using (\ref{4.9ae}))
\bminiG{4.11e}
\label{4.11ae}
3{\dot{a}^2 \over a^2 b^2} = \kappa \ T_{M 00}\ , 
\eeeq  
\beeq
\label{4.11be}
2{\ddot{a} \over ab^2} + {\dot{a}^2 \over a^2 b^2}  - 2
{\dot{a}\dot{b} \over ab^3} = \kappa \ T_{M , S}\ .
\emini

These equations are invariant under time-like diffeomorphisms that
allow the gauge $b(t) = 1$, after which they turn into the standard
FRW equations that enforce the conservation of the standard stress 
energy tensor.\par

Next we have analysed the system for arbitrary standard cosmological
constant,
\beq
\label{4.12e}
T_{M 00}\ =\ T_{M, S}\ =\ \Lambda\ ,
\eeq
and arbitrary dual cosmological constant,
\beq
\label{4.13e}
\widetilde{T}_{M, S}\ =\ \widetilde{\Lambda}\ .
\eeq

The system of equations can be reduced by eliminating $\widetilde{r}$
using (\ref{4.2e}), $\dot{\widetilde{s}}$ using (\ref{4.3be}), and
introducing
\beq\label{4.14e}
\overline{\ell}^{(1)}_c = \overline{\ell}^{(1)}\ \widetilde{a}^2 =
\mathrm{const.}
\eeq
that solves (\ref{4.4be}). Next, $r$ can be eliminated using 
(\ref{4.8ae}), and $b$ using (\ref{4.9be}). We are left with 4 equations
for $a$, $\widetilde{a}$, $s$ and ${\ell}^{(1)}$, where the maximal time
derivatives are $\dot{a}$, $\ddot{\widetilde{a}}$, $\dot{s}$ and 
$\ddot{\ell}^{(1)}$:

\bea
\label{4.15e}
\dot{a}&=& \frac{1}{2 \ell^{(1)}}
\left (\frac{sb}{\kappa}-a 
\dot{\ell}^{(1)}\right ) \nn \\
\ddot{\widetilde{a}}&=& \frac{1}{\widetilde{a}^2 \ell^{(1)}} 
\left(
\dot{r} a^2 \left(\frac{1}{2\kappa}+\frac{\overline{\ell}^{(1)}_c}
{\widetilde{a}^2}\right)
+\frac{a^3 b}{6}\Lambda\right) \nn \\
\dot{s}&=& \frac{\widetilde{a}^2}{a^2}\ddot{\widetilde{a}} \nn \\
\ddot{\ell}^{(1)}&=& - 
2\dot{r}\frac{a^2}{\widetilde{a}^3} \left (\frac{1}{\kappa}
+\frac{\overline{\ell}^{(1)}_c}{\widetilde{a}^2}\right ) 
-4\dot{\ell}^{(1)}
\frac{\dot{\widetilde{a}}}{\widetilde{a}} 
-2{\ell}^{(1)}\frac{\dot{\widetilde{a}}^2}
{\widetilde{a}^2} -\frac{2a^3b}{3\widetilde{a}^3} \Lambda
+\frac{1}{2} \widetilde{\Lambda}
\eea
where one has to replace
\bminiG{4.16e}
\label{4.16ae}
b =-\frac{\dot{a}}{r}+\frac{2\kappa \overline{\ell}^{(1)}_c}
{r\widetilde{a}^3} \left(
a\dot{\widetilde{a}} -\dot{a}\widetilde{a} \right)\ , 
\eeeq  
\beeq
\label{4.16be}
r = \pm \sqrt{s^2 +\frac{\kappa}{3}a^2\Lambda}\ .
\emini

This system can be brought into normal form (i.e. be solved for
$\dot{a}$, $\ddot{\widetilde{a}}$, $\dot{s}$ and  $\ddot{\ell}^{(1)}$
after the replacements (4.16)) which is suitable for analytic and
numerical stability analyses.\par

Remarkably we found for a wide range of initial conditions, and
\underline{arbitrary} cosmological constants $\Lambda$ and
$\widetilde{\Lambda}$ (and $\overline{\ell}^{(1)}_c$), an
asymptotically stable (constant) solution for the dual scale factor
$\widetilde{a}$:
\bea
\label{4.17e}
&&\widetilde{a}\ \to \ \widetilde{a}_0 \nn \\
&& a \ \to a_0/t\nn \\
&& s\ \to \ s_0 \nn \\
&& {\ell}^{(1)}\ \to t^2 \widetilde{\Lambda}/4
\eea
where the constants $\widetilde{a}_0$, $a_0$ and $s_0$ depend on the
initial values. Another stable solution is given by
\bea
\label{4.18e}
&&\widetilde{a}\ \to \ t\widetilde{a}_0 \nn \\
&& a \ \to a_0/t\nn \\
&& s\ \to \ s_0 \nn \\
&& {\ell}^{(1)}\ \to t^2 \widetilde{\Lambda}/24
\eea
that is separated from (\ref{4.17e}) mainly through the initial values
for $\dot{\widetilde{a}}$ and $\dot{\ell}^{(1)}$. Note that the
$\widetilde{\Lambda}$-dependent values of $\ell^{(1)}$ are assumed
dynamically.\par

Provided that we identify our known matter with the dual matter of the
model, these solutions come very close to a solution of the CCP, since
the dual scale factor remains asymptotically constant (or increases
just linearly in $t$) for arbitrary $\widetilde{\Lambda}$, without
fine tuned initial conditions.\par

Let us compare the above solutions to the equations of the previous
section 3, in particular to the points i) to v) just before eq. (3.1),
in order to investigate at which point the impact of 
$\widetilde{\Lambda}$ on the dual scale factor $\widetilde{a}$ gets
lost. First, the impact of $\widetilde{\Lambda}$ on  $L^{(1)\ \ c}_{\ \
ab}$ in the form of $\ell^{(1)}$ is evident. The crucial equation is
eq. (4.9b), where the impact of $\ell^{(1)} \sim t^2$ on $s$ (which
corresponds to torsion $\omega^T$) is cancelled through the decay of
$a(t) \sim 1/t$, up to subleading terms that allow for $s \to s_0$.
(Remarkably this does not imply that the ``original" universe is
contracting: from (4.16a) one obtains $b(t) \sim b_0/t^2$, hence $t$
does not correspond to the cosmological time in this universe. Its
cosmological time is rather given by $t' = -b_0/t$, in terms if which
the signs of all components of $e^a_{\ \mu}$ change and $a(t')$
increases as $|a(t')| \sim \frac{a_0}{b_0} t'$. Note furthermore that
increasing $|t'|$ corresponds to decreasing $|t|$, i.e. the relative
arrows of time are reversed.)

However, before this can be considered as a  fully acceptable solution
of the CCP, the following tasks have to be performed:

\noi a) dual matter has to be added in order to check, whether the
known part of the evolution of our universe can be reproduced in the
dual sector,

\noi b) a weak field expansion around such a solution has to be
performed in order to see whether the (linearized) 
Einstein equations for the dual metric do not deviate too much from its
standard form.\par

We have performed a preliminary analysis in this direction by adding a
cosmological perturbation $\Delta \widetilde{T}$ to $\widetilde{T} =
\widetilde{\Lambda}$. In the case of the validity of the standard
Einstein equations this perturbation should induce a perturbation
\beq
\label{4.19e}
2 \ddot{\widetilde{a}} = \widetilde{\kappa}\ \Delta \widetilde{T}
\eeq
of the dual scale factor $\widetilde{a}$ (cf. (4.11b)). The good news
is that, neglecting terms of relative order $\sim t^{-1}$, the induced
perturbation of $\widetilde{a}$ can indeed be written in the form
(4.19). The bad news is that in both cases $\widetilde{\kappa}$ is
time dependent as $\widetilde{\kappa} \sim t^{-1}$ in the case of
(4.17), even $\widetilde{\kappa} \sim t^{-3}$ in the case of (4.18). (A
similar problem persists for the ``original" universe, which differs
also from a de Sitter universe for arbitrary $\Lambda$, unless
$\widetilde{\Lambda}=0$ and fine tuned initial conditions
$s = \dot{\widetilde{a}} =\dot{\ell^{(1)}} =0$ are used, i.e. the CCP
seems to be solved also here: however, the $t^2$ dependent value for
$\ell^{(1)}$, plugged into the equations of motion (2.23), shows that
the effective gravitational coupling for the ``original" universe --
its dependence on the Lorentz vector $n^d$ -- is also time dependent.)
Such time dependent gravitational couplings seem to be in conflict
with the cosmological  standard model, hence further studies of
possible modifications of the model are required.

\mysection{Summary and outlook}
\textheight= 23 truecm
\hspace*{\parindent}  We have constructed an action including dual
gravitational fields, without adding new degrees of freedom. An amazing
feature of the model is that some kind of matter can be coupled to the
original gravitational fields (and ``see'' a space-time geometry
described by the original metric), whereas another kind of matter can
be coupled to the dual metric and, hence, propagate on a generally
different space-time geometry. In spite of the manifest breaking of
Lorentz symmetry in the action, the reactions of the two different
space-time geometries to the corresponding stress energy tensors
coincide with Einstein's gravity in both cases, to lowest order in a
weak field expansion around Minkowski space-time.\par

Beyond lowest order, this phenomenon persists only for the standard
gravi\-tational fields (in the absence of dual matter): Whereas
fluctuating standard gravitational fields imply fluctuating dual
gravitational fields from eqs. (\ref{2.14e}) and (\ref{2.19e}), a
vanishing $\widetilde{T}_{M a}^{\ \ \ b}$ allows for vanishing
$L^{(1)}$ and $L^{(2)}$ from (\ref{2.29e}) and (\ref{2.31e}), as a
consequence of which eqs. (\ref{2.23e}) and (\ref{2.27e}) turn into the
standard equations for the standard vierbein and connection. \par

The opposite statement is not true; an attempt to
integrate out the standard gravitational fields (even for $T_{M a}^{\
\ \ b}=0$) will generate non-local effective interactions for the dual
fields, that are moreover expected to break Lorentz symmetry manifestly
in the form of (positive) powers of $\partial_0^2/\sq$. Hence it is an
open question up to now, whether our world could be identified with the
dual sector of the model.\par

This question is not purely academic, since the cosmological evolution
of the dual sector differs dramatically from standard cosmology: A
cosmological constant $\widetilde{\Lambda}$ does not imply an
exponential increase (with cosmic time) of the dual scale factor
$\widetilde{a}$, but can be ``absorbed'' completely into a time
dependence of the original scale factor. As stated at the end of
section 4, such a solution of the cosmological constant problem
requires further investigations.\par

Furthermore, various ways to generalize the model could be studied:
\par

a) As stated near the end of section 2, the fixed Lorentz vector $n^a$
could be replaced by a field $n^a(x)$ that breaks Lorentz symmetry
spontaneously. The present approach to gravitational $S$-duality would
then be similar in spirit to the PST approach \cite{23r} (reviewed in
\cite{11r}), that has been applied to $d= 10/11$ supergravities.\par

b) In higher dimensional, e.g. 5 dimensional, space times the dual of
$R^{ab}_{\ \ \mu\nu}$ would be, after contracting the $5d$ epsilon
tensor with the Lorentz indices, a tensor $\widetilde{R}^{abc}_{\ \ \
\mu\nu}$ that can be interpreted as a field strength of a field
$\widetilde{D}^{abc}_{\ \ \ \mu}$ (antisymmetric in $[abc]$). As
before, this field could be fixed completely by an axial gauge
condition and the duality constraint contracted with $n^{\mu} = n^5$.
Subsequently the field strength of $\widetilde{\omega}^{ab}_{\ \ \mu}$
can be obtained from $n_c \widetilde{R}^{abc}_{\ \ \ \mu\nu}$, and
$\widetilde{e}^{a}_{\  \mu}$ from the torsion tensor contracted with
$n^a$ as before (imposing the axial gauge condition on all these
fields). A systematic procedure for arbitrarily high dimensional
space-times could be developed along these lines, and it would be
interesting to study the duals of brane universes in such models. (The
4d metric that is S-dual to 2-brane universes could be investigated
already in the present model).\par

c) A weak point of the present model is that, in spite of the
introduction of dual gravitational fields, we did not manage to make
additional symmetries (as of the Ehler's type \cite{24r}) manifest,
even after dimensional reduction along a coordinate along $n^{\mu}$.
The technical problem here is that the standard  $t$ independent
dimensional reduction ansatz for the metric does not coincide with our
axial gauge conditions on $\widetilde{e}^{a}_{\  \mu}$ and
$\widetilde{\omega}^{ab}_{\  \ \mu}$.\par

It may then be advisable to carry out the essential steps of the
concept presented here -- the introduction of dual gravitational fields
together with non-covariant gauge conditions and  partial
(non-covariant) duality constraints -- in terms of different variables
as non-linear realizations of gravity \cite{25r}, or to give up the
axial gauge condition on $\widetilde{e}^{a}_{\  \mu}$ (which would
complicate the analysis considerably, once $n^{\mu}$ can no longer be
assumed to be constant). \par

In view of the interesting properties of the present model we believe
that these various open questions merit corresponding studies.

\section*{Acknowledgment}

It is a pleasure to thank J. Mourad for comments and discussions.

\end{document}